\documentclass[preprint,nopacs,preprintnumbers,amsmath,amssymb]{revtex4}
\usepackage{graphicx}
\begin{document}

\title{ Polarizability and impurity screening for phosphorene }
\author{Po Hsin Shih$^{1}$, Thi Nga Do$^{2}$, Godfrey Gumbs$^{3,4}$ and  Dipendra Dahal$^3$}
\affiliation{$^{1}$Department of Physics, National Cheng Kung University, Taiwan 701\\
$^{2}$Institute of Physics, Academia Sinica, Taipei, Taiwan 115\\
$^{3}$Department of Physics and Astronomy, Hunter College of the City University of New York,\\
695 Park Avenue, New York, NY 10065, USA \\
 $^{4}$Donostia International Physics Center (DIPC),
P de Manuel Lardizabal, 4, 20018 San Sebastian, Basque Country, Spain
}

%
%
% \date{\today}

  \begin{abstract}

\medskip

Using a tight-binding Hamiltonian for phosphorene,  we have calculated the real part of the polarizability and the corresponding dielectric function, Re[$\epsilon(\bf{q},\omega)$], at zero temperature (T = 0) with free carrier density $10^{13}/ cm^{−2}$.
We present results showing Re[$\epsilon(\bf{q},\omega)$] in different directions of the transferred momentum $\bf{q}$. When $q$ is larger than a particular value which is twice the Fermi momentum $k_F$, Re[$\epsilon(\bf{q},\omega)$]  becomes strongly dependent on the direction of $\bf{q}$. We also discuss the case at room temperature (T = 300K). These results which are similar to those previously reported by other authors are then employed to determine the static shielding of an impurity in the vicinity of phosphorene.

  \end{abstract}

\vskip 0.2in


\medskip
\par
  \maketitle

%\maketitle

\section{Introduction}
\label{int}

Emerging phenomena in physics and quantum information technology have relied extensively on the collective properties of low-dimensional materials such as two-dimensional (2D) and few-layer structures with nanoscale thickness.  There, the Coulomb and/or atomic interactions play a crucial role in these  complexes which include doped as well as undoped graphene \cite{GR1,GR2,GR3},
silicene \cite{Si4,Si5}, phosphorene \cite{BP9,BP10},
germanene \cite{Ge6,Ge7}, antimonene \cite{Sb11,Sb12}, tinene \cite{Ti8},
bismuthene \cite{Bi13,Bi14,Bi15,Bi16,Bi17,Bi18} and most recently the
2D pseudospin-1 $\alpha-T_3$ lattice  \cite{Ejnicol}.  Of these which have been successfully synthesized by various experimental techniques and which have been extensively investigated by various experimental techniques,  few-layer black phosphorus (phosphorene) or BP has been produced by using mechanical cleavage
\cite{PLi2014,PLiu2014}, liquid exfoliation \cite{PBrent2014,BP10,Pkang2015}, and mineralizer-assisted short-way transport reaction \cite{PLange2007,PNilges2008,PKopf2014}.

\medskip
\par
Unlike graphene, phosphorus inherently has an appreciable  band gap. The observed photoluminescence peak of single-layer phosphorus  in the visible optical range shows that its band gap is larger than that for bulk. Furthermore, BP has a middle energy gap ($~\sim 1.5-2$ eV) at the $\Gamma $ point, thereby being quite different from the narrow or zero gaps of group-IV systems. Specifically, experimental measurements have shown that the BP-based field effect transistor has an on/off ratio of 105 and a carrier mobility at room temperature as large as 103 cm$^{2}/$Vs. We note that BP is expected to play an important role in the next-generation of electronic devices \cite{PLi2014,PLiu2014}. Phosphorene exhibits a puckered structure related to the $sp^{3}$ hybridization of ($3s,3p_{x},3p_{y},3p_{z}$) orbitals. The deformed
hexagonal lattice of monolayer BP has four atoms \cite{PRudenko}, while the
group-IV honeycomb lattice includes two atoms. The low-lying energy dispersions, which are dominated by $3p_{z}$ orbitals, can be described by a four-band model with complicated multi-hopping integrals \cite{PRudenko}. The low-lying energy bands are highly anisotropic, e.g., the linear and parabolic dispersions near the Fermi energy $E_{F}$, respectively, along the $\widehat{k_{x}}$ and $\widehat{k_{y}}$ directions. The anisotropic behaviors are further reflected in other
physical properties, as verified by recent measurements on optical and excitonic spectra
\cite{Oleg} as well as transport properties \cite{PLi2014,PLow}.

\medskip
\par
In this work, we have examined the anisotropic behavior of the static polarizability and
shielded potential of an impurity for BP. The calculations for the polarizability were executed at T=0K and room temperature (T=300K).  We treat the buckled BP structure as a 2D sheet in our formalism. Consequently, we present an algebraic expression for the surface response function of a pair of 2D layers with arbitrary separation and which are embedded in dielectric media. We then adapt this result to the case when the layer separation is very small to model a free-standing buckled BP structure.

\medskip
\par

The outline of the rest of our presentation is as follows. In Sec.\ \ref{sec2},
we  present the surface response function for a pair of 2D layers embedded in
background dielectric media. We then simplify this result for a pair of planar sheets which are infinitesimally close to each other and use this for  buckled BP.
The tight-binding model Hamiltonian for BP is presented in Sec. \ref{sec3}. This is employed in our calculations of the energy bands and eigenfunctions. Section  \ref{sec4}. is devoted to the calculation of the polarizability and dielectric function of BP
showing its temperature dependence and their anisotropic properties as a consequence of its band structure. Impurity shielding by BP is discussed in Sec. \ref{sec5} and we summarize our important results in Sec. \ref{sec6}.

\section{Surface response function for a pair of 2D layers}
\label{sec2}

\medskip
\par

The external potential will give rise to an induced potential which, outside the
structure,  can be written as

\begin{equation}
\phi_{ind}({\bf r}_\parallel,t)=-\int \frac{d^2{\bf q}}{(2\pi)^2}\int_{-\infty}^\infty d\omega\
\tilde{\phi}_{ext}\left( q,\omega\right)  e^{i({\bf q}\cdot {\bf r}_\parallel-\omega t)}
g({\bf q},\omega)e^{-q z} \ .
\end{equation}
This equation defines the surface response function $g({\bf q},\omega)$.  It has been implicitly assumed that
the external potential $\phi_{ext}$ is so weak that the medium responds linearly to it.

\medskip
\par

The quantity  Im[g($\bf{q}, \omega)$]  can be identified with the power absorption in the
structure  due to electron excitation induced by the external potential.
The total potential in the vicinity of the surface ($z\approx 0$),  is given by

\begin{equation}
\phi ({\bf r}_\parallel,t) =     \int \frac{d^2{\bf q}}{(2\pi)^2}\int_{-\infty}^\infty d\omega\
\left( e^{q z}-g({\bf q},\omega) e^{-q z} \right)
 e^{i({\bf q}\cdot {\bf r}_\parallel  -\omega t)}
\tilde{\phi}_{ext}\left( {\bf q},\omega\right)
 \label{eq:8}
\end{equation}
which takes account of nonlocal screening of the external potential.

\subsection{Model for phosphorene layer}

\begin{figure}
\centering
\includegraphics[width=.55\textwidth]{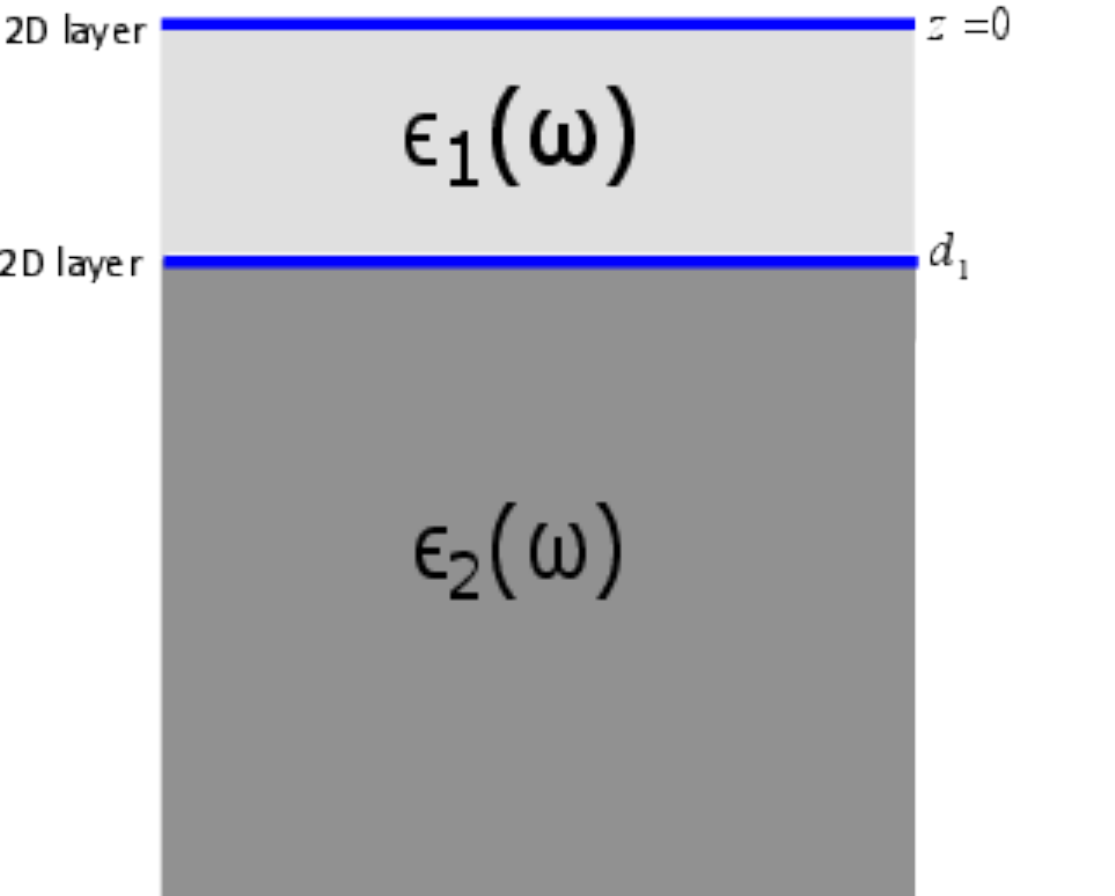}
\caption{(Color online)  Schematic illustration of a hybrid structure consisting of a pair of 2D layers
separated by distance $d_1$.}
\label{FIG:1}
\end{figure}

In this section, we present the surface repsonse function we calculated for a structure which consists of a pair of 2D layers in contact with a dielectric medium, as shown in Figure\ \ref{FIG:1}
. One of the 2D layers is at the top and the other is
encapsulated by materials with dielectric constants $\epsilon_1(\omega)$, with thickness $d_1$, and $\epsilon_2(\omega)$, of semi-infinite thickness.
Calculation shows that the surface response function is given by \cite{Book,PRB2018}

\begin{equation}
 g({\bf q} ,\omega)=\frac{{\cal N}({\bf q},\omega)}{{\cal D}({\bf q},\omega)} \ ,
\end{equation}
where

\begin{eqnarray}
{\cal N}({\bf q},\omega) &\equiv &
 e^{2 d_1 q } \{q  \epsilon_0 (\epsilon_1(\omega)-1)-\chi_1({\bf q},\omega)\} \{q
\epsilon_0 (\epsilon_1(\omega)+\epsilon_2(\omega))-\chi_2({\bf q},\omega)\}
\nonumber\\
&-&\{q  \epsilon_0 (\epsilon_1(\omega)+1)+\chi_1({\bf q},\omega)\} \{q
\epsilon_0 (\epsilon_1(\omega)-\epsilon_2(\omega))+\chi_2({\bf q},\omega)\} \ ,
\end{eqnarray}
and

\begin{eqnarray}
{\cal D}({\bf q},\omega) &\equiv&
e^{2 d_1 q } \{q  \epsilon_0 (\epsilon_1(\omega)+1)-
\chi_1({\bf q},\omega)\} \{q  \epsilon_0 (\epsilon_1(\omega)+\epsilon_2(\omega))-\chi_2({\bf q},\omega)\}
\nonumber\\
&-&\{q  \epsilon_0 (\epsilon_1(\omega)-1)+\chi_1({\bf q},\omega)\}
 \{q  \epsilon_0 (\epsilon_1(\omega)-\epsilon_2(\omega))+\chi_2({\bf q},\omega)\}  \ .
\end{eqnarray}
In this notation, ${\bf q}$ is the in-plane wave vector, $\omega$ is the frequency
and $\chi_1({\bf q},\omega)$ and $\chi_2({\bf q},\omega)$ are the 2D layer susceptibilities.

\begin{figure}
\centering
\includegraphics[width=.55\textwidth]{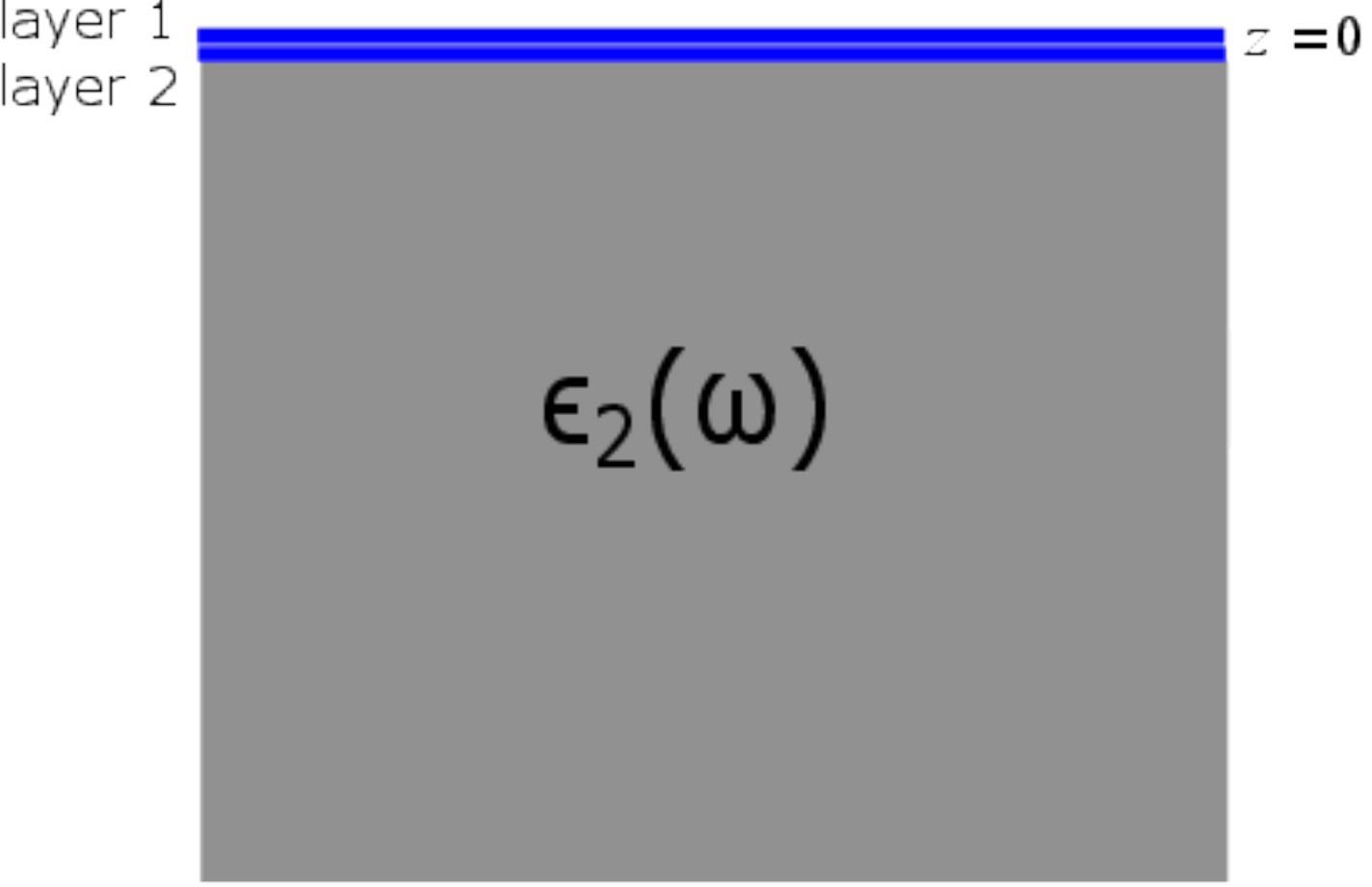}
\caption{(Color online)  Schematic representation of a structure consisting of  a pair of 2D
layers which are infinitesimally close. There is vacuum above the layers and a dielectric below.}
\label{FIG:2}
\end{figure}

When we take the limit $d_1\to 0$, i.e., the  separation between the two layer is small,
the $\epsilon_1$ drops out and we have the following result for the surface response
function corresponding to the structure in Figure\ \ref{FIG:2}

\begin{equation}
g({\bf q},\omega)=
1-\frac{1}{\frac{1+\epsilon_2(\omega)}{2}-\frac{\chi_1({\bf q},\omega)+\chi_2({\bf q},\omega)}{2 q \epsilon_0}}
\ .
\label{IMP}
\end{equation}
Here, the dispersion equation which is given by the zeros of the denominator $\epsilon({\bf q},\omega)$
of the second term  is expressed in terms of the 'average' susceptibility for the two layers. Clearly, this
dispersion equation is that for a 2D layer of the Stern form where we make the identification
$\chi\to  e^2\Pi^{(0)}$ in terms of the polarizability. This  result in Eq.\ (\ref{IMP}) clearly illustrates
that for the buckled BP structure shown in Figure\ \ref{FIG:3}, the dielectric function can be treated as that
for a single layer whose susceptibility arises from a {\em combination\/} of two rows of atoms making up the
layer.  Our calculation can easily be generalized to the case when the monolayer is embedded above and below by
 the same thick dielectric material (dielectric constant $\epsilon_b$) which corresponds to the free-standing
situation which we consider below.  For this, we have $\epsilon({\bf q},\omega)=\epsilon_b-
e^2/(2\epsilon_0 q)\Pi^{(0)}({\bf q},\omega)$, expressed in terms of the 2D layer polarizability
$\Pi^{(0)}({\bf q},\omega)$.

\medskip
\par

\section{Model Hamiltonian}
\label{sec3}

\begin{figure}[h]
\centering
{\includegraphics[height=9cm]{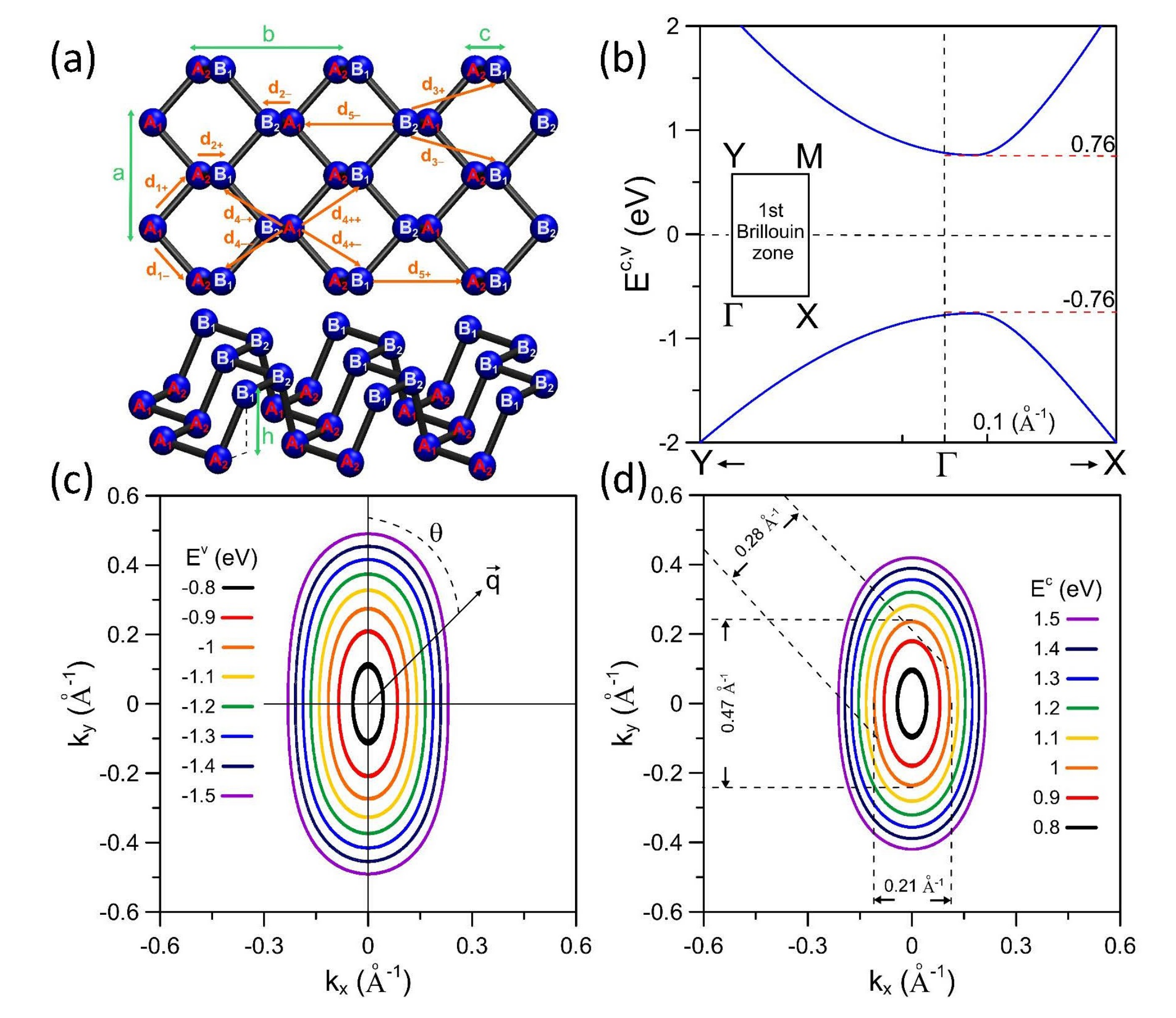}}
\caption{(Color online) The (a) top view and side view of crystal structure for BP and (b)
its band structure. The constant-energy diagrams are presented for (c) valence and (d) conduction  bands.
The values of 2$k_F$ for different $\theta$ are given in (d).}
\label{FIG:3}
\end{figure}

Phosphorene is treated as a single layer of phosphorus atoms arranged in a puckered orthorhombic lattice, as shown in Fig. \ref{FIG:3}(a). It contains two atomic layers of A and B atoms and two kinds of bonds for in-plane and inter-plane P-P connections with different bond lengths. The low-lying electronic structure can be described by a tight-binding Hamiltonian, which is a 4 $\times$ 4 matrix within the basis ($A_1$, $A_2$, $B_1$, $B_2$), of the form
\[
  \begin{bmatrix}
    0 & T_1 +T_3^* & T_4 & T_{2-} + T_{5-}^*  \\
    T_1^* +T_3 & 0 & T_{2+} + T_{5+}^* & T_4 \\
    T_4 & T_{2+}^* + T_{5+} & 0 & T_1 +T_3^* \\
    T_{2-}^* + T_{5-} & T_4 & T_1^* +T_3 & 0
  \end{bmatrix}.
\]
Here, we consider up to five nearest atomic interactions through five independent terms of $T_i$ with $i = 1, 2, 3, 4, 5$. These terms are given by the following expressions.

\begin{eqnarray}
\left\{\begin{array}{ll}
T_1 = t_1 e^{i   {\bf k} \cdot (  {\bf d}_{1+} +   {\bf d}_{1-})} \\
T_{2\pm} = t_2 e^{i   {\bf k}\cdot {\bf d}_{2\pm}} \\
T_3 = t_3 e^{i  {\bf k} \cdot (  {\bf d}_{3+} +  {\bf d}_{3-})}  \\
T_4 = t_4 e^{i  {\bf k} \cdot(  {\bf d}_{4++} + \vec {d_{4+-}}+ \vec {d_{4-+}}+   {\bf d}_{4--})}  \\
T_{5\pm} = t_5 e^{i  {\bf k} \cdot {\bf d}_{5\pm}}.
\end{array} \right.
\end{eqnarray}
In this notation, $t_m$ ($m = 1, 2, 3, 4, 5$) are the hopping integrals, corresponding to the atomic interactions. They have been optimized as ($t_1 = -1.220$, $t_2 = 3.665$, $t_3=-0.205$, $t_4 =-0.105$, $t_5=-0.055$) in order to reproduce the energy bands obtained by the density functional theory (DFT) calculations \cite{dft1, dft2, tbm1}. Also,  $\vec {d_{m\pm}}$ are the vectors connecting the lattice sites which can be written as

\begin{eqnarray}
\left\{\begin{array}{ll}
d_{1\pm} = (b/2 - c, \pm a/2, 0) \\
d_{2\pm} =  (\pm c, 0, h)\\
d_{3\pm} =   (b/2 +c, \pm a/2, 0)\\
d_{4\pm} =   (\pm b/2, \pm a/2, h)\\
d_{5\pm} = \{\pm(b-c), 0, -h\},
\end{array} \right.
\end{eqnarray}
where $a=3.314 \AA$, $b=4.376 \AA$, $c=0.705 \AA$, and $h = 2.131 \AA$ are the distances between
the BP atoms  \cite{tbm2, tbm3}, as illustrated in Figure\ \ref{FIG:3}(a).

\medskip
\par

The valence and conduction energy bands present strong anisotropic behaviors, as illustrated by the energy bands in Fig. \ref{FIG:3}(b) and  the constant-energy loops in Figs. \ref{FIG:3}(c) and \ref{FIG:3}(d). As a result, the polarizability and dielectric function are shown to be strongly dependent on the direction of the transferred momentum $  {\bf q}$.

\begin{figure}[h]
\centering
\includegraphics[width=.75\textwidth]{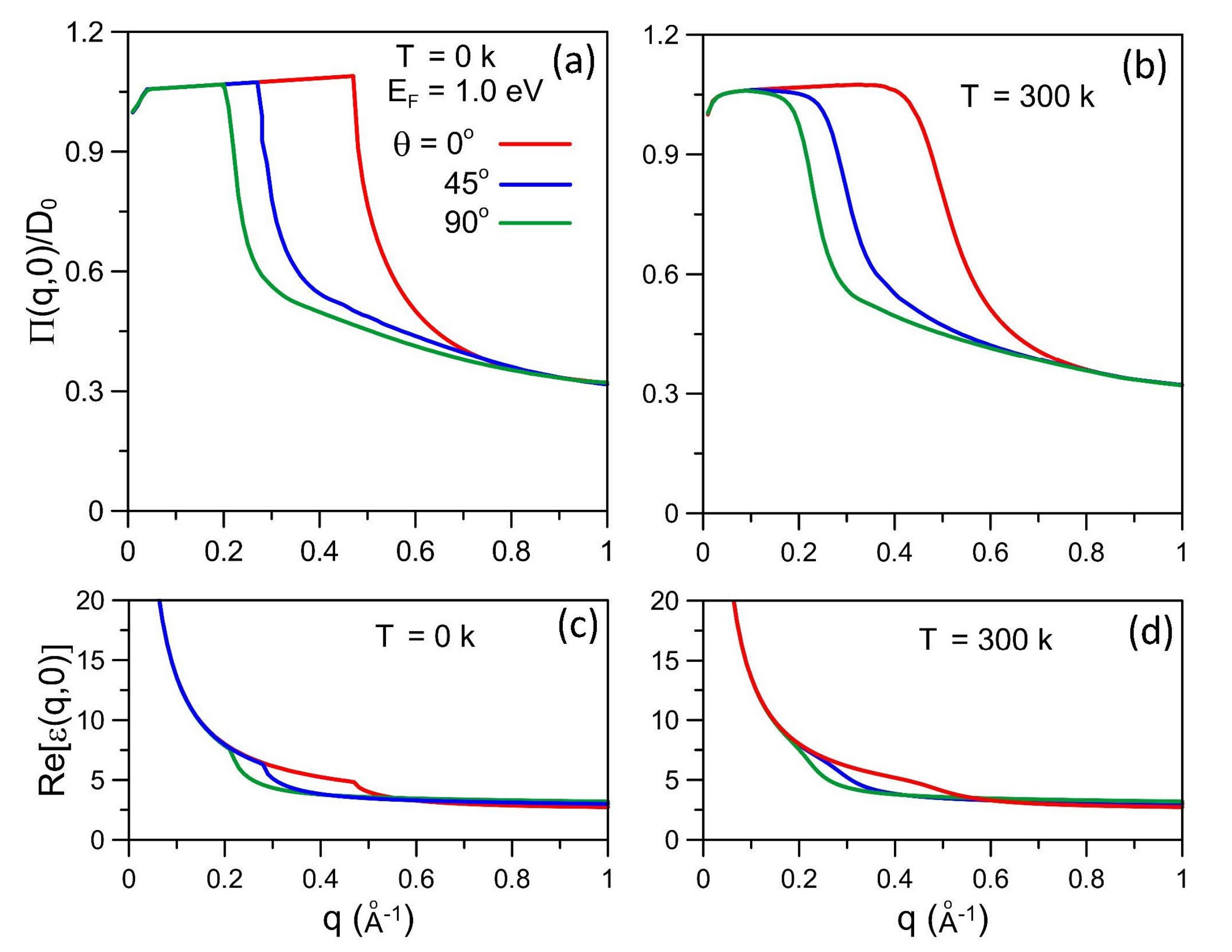}
\caption{(Color online)  The static polarizability for  BP as a function of wave vector for different directions of the transferred momentum $  {\bf q}$ at (a) zero and (b) room temperatures. Plots (c) and (d) correspond to the static dielectric function of BP at T = 0   and T = 300K, respectively.}
\label{FIG:4}
\end{figure}

\section{Dielectric Function}
\label{sec4}

When monolayer BP is perturbed by an external time-dependent Coulomb potential, all the valence and
conduction electrons will screen this field and therefore create the charge redistribution.
The effective potential between two charges is the sum of the external potential and the induced potential
due to screening charges. The dynamical dielectric function, within the random-phase approximation (RPA), is given by \cite{diele}

\begin{equation}
\begin{aligned}
\epsilon( {\bf q},\omega)=\epsilon_{b}-V_{q}\displaystyle\sum_{s,s'=\alpha,\beta}\displaystyle\sum_{h,h'=c,v}\int_{1stBZ}\frac{\mathrm{d}k_{x}\mathrm{d}k_{y}}{(2\pi)^{2}}\vert \langle s^\prime;h^\prime; {\bf k}+ {\bf q} \vert \mathrm{e}^{\mathrm{i} {\bf q}\cdot {\bf r}_\parallel}\vert s;h; {\bf k}\rangle\vert^{2} \\
\times \frac{f(E^{s^\prime,h^\prime}( {\bf k}+ {\bf q}))-f(E^{s,h}( {\bf k}))}{E^{s',h'}(  {\bf k}+ {\bf q})
-E ^{s,h}( {\bf k})-(\omega+\mathrm{i}\Gamma)} \  .
\end{aligned}
\end{equation}Here, the $\pi$-electronic excitations are described in terms of the transferred momentum $ {\bf q}$
and the excitation frequency $\omega$.
$\epsilon_b=2.4$ the background dielectric constant, $V_{q} =$ 2$\pi\mathrm{e}^2/(\varepsilon_s q)$ the 2D Fourier transform of the bare Coulomb potential energy ($\varepsilon_s=4\pi\epsilon_0$), and $\Gamma$ the energy width due to various de-excitation mechanisms. $f(E) = 1/\{1+\mathrm{exp}[(E-\mu)k_BT]\}$ the Fermi-Dirac distribution in which $k_B$ is the Boltzmann constant and $\mu$ the chemical potential corresponding to the highest occupied state energy (middle energy of band gap) in the (semiconducting) metallic systems at T=0.

\medskip
\par

Figures \ref{FIG:4}(a) and \ref{FIG:4}(b) show the directional/$\theta$-dependence of the static polarization function $\Pi^{(0)} (0,{\bf q})$, in which $\theta$  defines the angle between the direction of ${\bf q}$ and the unit vector $\hat{k}_y$. For arbitrary $\theta$, the polarization function at lower ($q \leq$ 0.2 ($1/\AA$)) and higher ($q \geq$ 0.7 ($1/\AA$)) transferred momentum remains unchanged. In general, $\Pi^{(0)}  (0,{\bf q})$ falls off rapidly beyond a critical value of $q$ (2$k_F$)   which depends on $\theta$. For increasing $\theta$ from 0 to $90^{\circ}$, the specific values are getting larger, as shown in Fig. \ \ref{FIG:4}(a). This means that the polarizability is stronger for 0.2 $\leq q \leq$ 0.7 ($1/\AA$). The main features of the polarizability for BP are quite similar to those for the 2D electron gas, but different with those for graphene. Temperature has an effect on the polarization function which is demonstrated in Fig. \ref{FIG:4}(b). At  room temperature, $\Pi^{(0)}(0,{\bf q})$ exhibits a shoulder-like structure near the critical values of $q$ instead of step-like structure at T = 0.

\par
\medskip
\par

Plots of the static dielectric function of BP for various values of $\theta$ are presented in Figs. \ref{FIG:4}(c) and \ref{FIG:4}(d) at zero and room temperatures, respectively. In the range of 0.2 $\leq q \leq$ 0.5 ($1/\AA$), there is a clear dependence of the dielectric function on the direction of the transferred momentum ${\bf q}$. The Re $\epsilon(0,q)$ is higher with the growth of $\theta$. The introduction of  finite temperature smoothens the $q$-dependent Re $\epsilon(0,q)$, as shown in Fig. \ref{FIG:4}(d) for T = 300K.

\vskip 0.3in

 \section{Impurity shielding}

 \label{sec5}

Starting with Eq.\ (\ref{eq:8}), we obtain the static screening of the potential on the surface at $z=0$ due to an impurity with charge $Z_0^\ast e$ located at distance $z_0$ above the surface of BP as

\begin{eqnarray}
\phi ({\bf r}_\parallel,\omega=0) &=&
  \frac{Z_0^\ast e}{2\pi \epsilon_0} \int_0^\infty  dq \int_{0}^{2\pi} d\theta\
e^{iq r\cos\theta}
\left[ 1-g({\bf q},\omega=0)  \right] e^{-q z_0}\
\nonumber\\
&=&
  \frac{Z_0^\ast e}{2\pi \epsilon_0} \int_0^\infty  dq \int_{0}^{2\pi} d\theta\
\frac{e^{iq r\cos\theta-q z_0}}{ \epsilon({\bf q},\omega=0) } \  .
\label{eq:8scr}
\end{eqnarray}

\begin{figure}
\centering
\includegraphics[width=.75\textwidth]{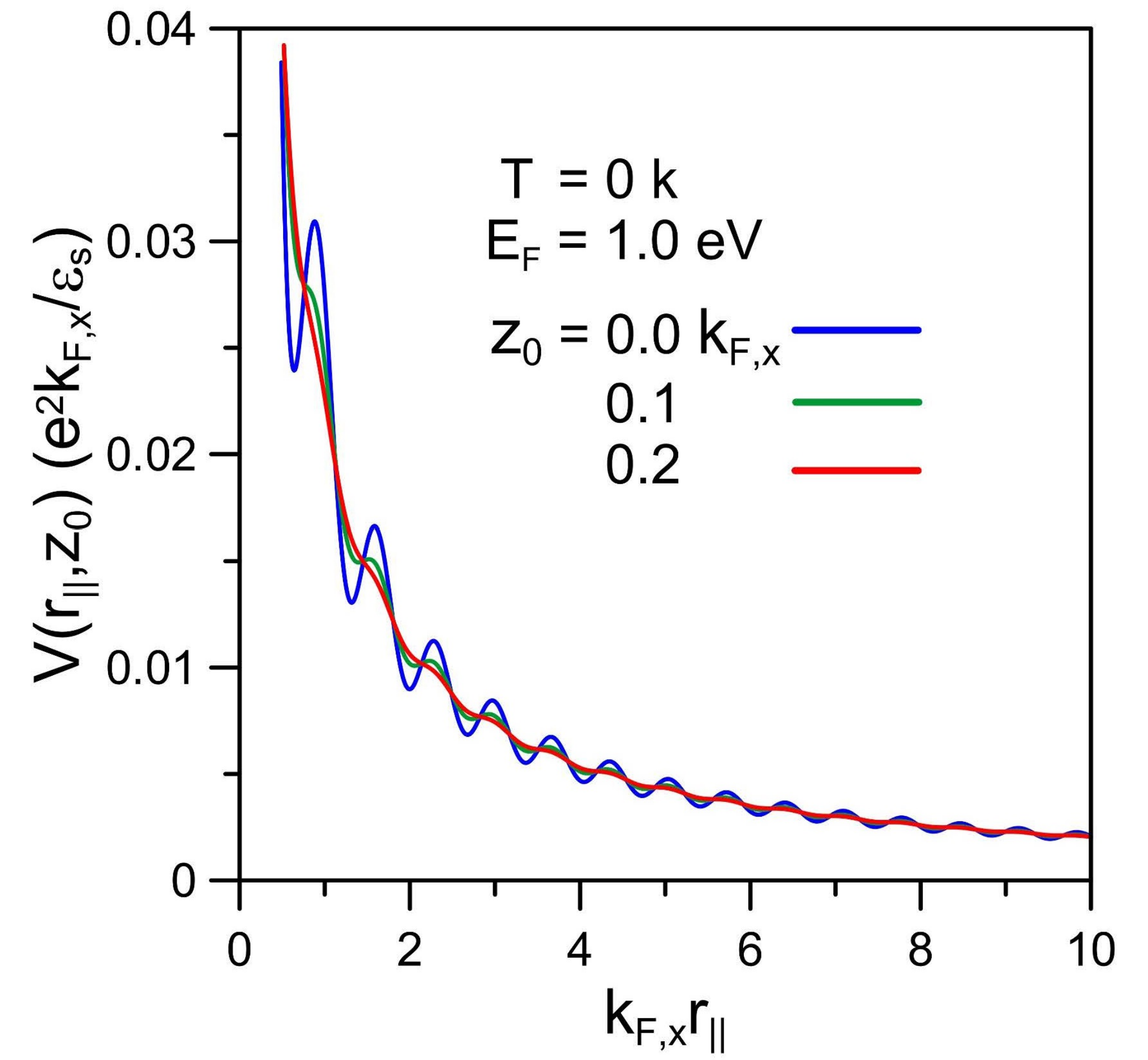}
\caption{(Color online) The screened impurity potential in units of $e^2 k_{F,x}/(\epsilon_s)$ is
plotted as a function of $k_{F,x}r_{||}$ for the chosen parameters in the figure.}
\label{FIG:5}
\end{figure}

\medskip
\par

By employing the generalized form of Eq. (\ref{IMP}) for free-standing BP in Eq.\ (\ref{eq:8scr}), we have computed the screened impurity potential.  The screened potentials for various $z_0$'s are shown in Fig. \  \ref{FIG:5} at zero temperature and Fermi energy $E_F$ = 1.0 eV. There exist Friedel oscillations for sufficiently small $z_0$. Such oscillations might be smeared out for larger $z_0$, e.g., the green and red curves. It is noticed that for $E_F$ = 1.0 eV, the room temperature of 300K which is much smaller than the Fermi temperature (10000K) does not have significant effect on the screened potential. Apparently, $V (r_{\parallel}, z_0)$ at T = 0 and T = 300K (not shown) are almost equivalent.

\medskip
\par

\section{Concluding Remarks and Summary}
\label{sec6}

 The energy band structure of BP, calculated using the tight-binding method, is anisotropic and so are
its polarizability, dielectric function and  screened potential.  To illustrate these
facts, we have presented numerical results for the polarizability in the $x$ and $y$ directions for a range of doping concentrations. The Re[$\epsilon({\bf q},\omega=0)$] of the static dielectric function for BP also reveals some interesting characteristics.
At zero temperature (T = 0) and with free carrier density  corresponding to chosen Fermi energy $E_F$, we have presented numerical results for Re[$\epsilon({\bf q},\omega=0)$]  in different directions of the transferred momentum ${\bf q}$. When $q$ is larger than a critical value which is twice the Fermi momentum $k_F$, our calculations show that Re[$\epsilon({\bf q},\omega=0)$]  becomes substantially dependent on the direction of ${\bf q}$. We also discuss the case at room temperature (T = 300K).  These results  are in agreement with those reported by other authors.  We employ our data to determine the static shielding of an impurity in the vicinity of phosphorene.
\medskip

\noindent
{\bf Conflict of interest}\\
All the authors declare that they have no conflict of interest.

\acknowledgments
G.G. would like to acknowledge the support from the Air Force
Research Laboratory (AFRL) through Grant \#12530960 .

\noindent
{\large {\bf References}}

\end{document}